\documentclass{article}

\usepackage{arxiv}
\usepackage{amsmath,amssymb,amsfonts}

\usepackage[utf8]{inputenc} 
\usepackage[T1]{fontenc}    
\usepackage{hyperref}       
\usepackage{url}            
\usepackage{booktabs}       
\usepackage{amsfonts}       
\usepackage{nicefrac}       
\usepackage{microtype}      

\usepackage{graphicx}
\usepackage{natbib}
\usepackage{doi}

\usepackage{amsmath}
\usepackage{amssymb}
\usepackage{algorithmic}
\usepackage{textcomp}
\usepackage{xcolor}
\usepackage{color}
\usepackage{diagbox}
\usepackage[ruled,vlined]{algorithm2e}

\usepackage{amsthm}
\usepackage[font=small]{caption}
\usepackage{subfigure}
\usepackage{graphics} 
\usepackage{epsfig} 
\usepackage{array}
\usepackage{url}
\usepackage{epstopdf}
\usepackage{enumerate}
\usepackage{paralist}
\AppendGraphicsExtensions{.tif}
\theoremstyle{remark}

\newtheorem{lemma}{Lemma}
\newtheorem{theorem}{Theorem}

\newtheorem{assumption}{Assumption}
\newtheorem{remark}{Remark}
\newtheorem{example}{Example}
\newtheorem{problem}{Problem}
\newtheorem{definition}{Definition}

\DeclareMathOperator*{\argmin}{arg\,min}

\title{Distributed Control-Estimation Synthesis for Stochastic Multi-Agent Systems via Virtual Interaction between Non-neighboring Agents}

\author{ \href{https://orcid.org/0000-0003-4325-3262}{\includegraphics[scale=0.06]{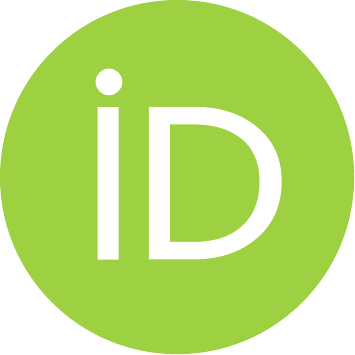}\hspace{1mm}Hojin Lee}\thanks{© 2021 IEEE. Personal use of this material is permitted. Permission from IEEE must be
obtained for all other uses, in any current or future media, including
reprinting/republishing this material for advertising or promotional purposes, creating new
collective works, for resale or redistribution to servers or lists, or reuse of any copyrighted
component of this work in other works. Citation information: DOI 10.1109/LCSYS.2021.3086848, IEEE Control Systems Letters.
This work was supported by the National Research Foundation of Korea(NRF) grants funded by the Korea government(MSIT) (No. 2020R1A5A8018822 and No. 2020R1C1C1007323} \\
	Department of Mechanical Engineering\\
	Ulsan National Institute of Science and Technology\\
	Ulsan, 44919 Repulic of Korea\\
	\texttt{hojinlee@unist.ac.kr} \\
	\And
	\href{https://orcid.org/0000-0002-6084-8236}{\includegraphics[scale=0.06]{orcid.pdf}\hspace{1mm}Cheolhyeon Kwon} \\
	Department of Mechanical Engineering\\
    Ulsan National Institute of Science and Technology\\
	Ulsan, 44919 Repulic of Korea\\
	\texttt{ kwonc@unist.ac.kr} \\
}

\date{}

\renewcommand{\headeright}{ }
\renewcommand{\undertitle}{ }
\renewcommand{\shorttitle}{ }

\hypersetup{
pdftitle={arXiv:2106.00961},
pdfsubject={MAS},
pdfauthor={Hojin Lee, Cheolhyeon Kwon},
pdfkeywords={Distributed Control},
}

\begin{document}
\maketitle

\begin{abstract}
This paper considers the optimal distributed control problem for a linear stochastic multi-agent system (MAS). Due to the distributed nature of MAS network, the information available to an individual agent is limited to its vicinity. From the entire MAS aspect, this imposes the structural constraint on the control law, making the optimal control law computationally intractable. This paper attempts to relax such a structural constraint by expanding the neighboring information for each agent to the entire MAS, enabled by the distributed estimation algorithm embedded in each agent. By exploiting the estimated information, each agent is not limited to interact with its neighborhood but further establishing the `virtual interactions' with the non-neighboring agents. Then the optimal distributed MAS control problem is cast as a synthesized control-estimation problem. An iterative optimization procedure is developed to find the control-estimation law, minimizing the global objective cost of MAS.
\end{abstract}


\section{Introduction}\label{sec:introduction}
Distributed control within a cooperative multi-agent system (MAS) is the key enabling technology for different networked dynamical systems \citep{shi2020survey,li2019fully,zhu2020sampled,morita2015multiagent}. Notwithstanding diverse distributed control strategies, their optimality is one of the stumbling blocks due to individual agents’ limited information. In particular, finding the optimal distributed control with network topological constraint is a well-known NP-hard problem \citep{gupta2005sub}. To ease this problem, some former studies have focused on a specific form of objective function along with certain MAS network topology conditions under which the optimal distributed control laws can be designed \citep{ma2015lqr}. More particularly, different techniques have been investigated to design suboptimal distributed control laws for different MAS cooperative tasks \citep{gupta2005sub,nguyen2015sub,jiao2019suboptimality}.
In this paper, a new avenue for accomplishing the optimal distributed control of MAS is presented while not requiring the restricted form of the objective function, nor the network topology. The key idea is to expand the available information for each agent by employing the distributed estimation algorithm, and use the expanded information to relax the network topological constraint in a tractable manner. In a nutshell, the main contributions are the following.
\begin{enumerate}
  \item A synthesized distributed control-estimation framework is proposed based on the authors' preliminarily developed distributed estimation algorithm \citep{kwon2018sensing}. The newly proposed framework enables the interactions between non-neighboring agents, namely \emph{virtual interactions}.
  \item With the aid of virtual interactions, a design procedure that solves for the optimal distributed control law of the stochastic MAS over a finite time horizon is developed, which was originally an intractable non-convex problem due to the network topological constraint.
\end{enumerate}

\section{Problem Formulation}\label{sc:problem_formulate}
\subsection{Dynamical Model of Stochastic MAS}
Consider a stochastic linear multi-agent dynamical system including $N$ homogeneous agents whose dynamics is given by:
  \begin{equation}\label{eq:state_equation}
x_{i}(t+1) = Ax_{i}(t)+Bu_{i}(t)+w_{i}(t),\ \ \ \  \forall i \in \{1,\cdots,N\}
\end{equation} 
where $x_{i}(t) \in \mathbb{R}^{n}$ and $u_{i}(t) \in \mathbb{R}^{p}$ are the state and the control input of the $i^{th}$ agent, respectively. $w_{i}(t)$ is a disturbance imposed on the $i^{th}$ agent, assumed to follow zero-mean white Gaussian distribution with covariance $\mathrm{\Theta}_{i}(t) \succ 0 $. $t \in \mathbb{Z}_{+}= \{0,1,\cdots \}$ indicates a discrete-time index. $A, B$ are the system matrices with appropriate dimensions, and are assumed to satisfy the controllability condition. Accordingly, the entire MAS dynamics can be written as \begin{equation}\label{eq:mas_dynamic}
x(t+1) = \tilde{A}x(t)+\tilde{B}u(t)+\tilde{w}(t)
\end{equation}
\begin{equation*}
\begin{split}
    \tilde{A} & = \left(I_{N}\otimes A\right), \ \tilde{B} = \left(I_{N}\otimes B\right) \\ 
x(t) &=\begin{bmatrix}x_{1}^{\mathrm{T}}(t) \cdots x_{N}^{\mathrm{T}}(t)\end{bmatrix}^{\mathrm{T}},\
 u(t) =\begin{bmatrix}u_{1}^{\mathrm{T}}(t) \cdots u_{N}^{\mathrm{T}}(t)\end{bmatrix}^{\mathrm{T}}\\
\tilde{w} & = \begin{bmatrix}w_{1}^{\mathrm{T}}(t) \cdots w_{N}^{\mathrm{T}}(t)\end{bmatrix}^{\mathrm{T}}
\end{split}    
\end{equation*}
where $\otimes$ is the Kronecker product between matrices. The interactions between agents are rendered by inter-agent network topology, described by a graph model $\mathcal{G}$ consisting of a node set $\mathcal{V}=\{1,\ 2\ .\ .\ ,\ N\}$ indexing each agent and an edge set $\mathcal{E}\subseteq \mathcal{V}\times \mathcal{V}$ indicating the network connectivity between the agents. Each edge $(i,j) \in \mathcal{E}$ denotes that the node $i$ can acquire the state information of the node $j$. An adjacency matrix $\mathcal{A}=[a_{ĳ}] \in \mathbb{R}^{N\times N}$ can express the network connectivity of the graph model, where its element $a_{ij}=1$ if $(i,j)\in \mathcal{E}$, and $a_{ij}=0$ otherwise. A degree matrix is defined as $\mathcal{D} = diag(d_{1}\cdots d_{N})$ where $d_{i}=\sum_{\substack{j}}a_{ij}$ is (weighted) degree of node $i$. The Laplacian matrix $\mathcal{L}$, given by $\mathcal{L} = \mathcal{D} - \mathcal{A}$, is useful for analysis of the network topology. The set of agents whose state information is available to the $i^{th}$ agent, i.e., the neighborhood of the $i^{th}$ agent, is expressed as $\Omega_i$, and its cardinality is expressed as $|{\Omega_i}|$. Based on the given network topology, the noisy measurement of neighborhood states $\{x_{j}(t)|j \in \mathcal{V}\}$ from the $i^{th}$ agent's perspective can be represented as follows \citep{kwon2018sensing}:
\begin{equation}\label{eq:measurement}
z_{ij}(t) = c_{ij}\left(x_{j}(t)+v_{ij}(t)\right), \ \ \ \ \ \forall j \in \mathcal{V}
\end{equation}
where $c_{ij}$ indicates the availability of the measurement of the $j^{th}$ agent's state from the $i^{th}$ agent such that $c_{ij} = 1$ when $j\in\Omega_{i}$, and $c_{ij}=0$ otherwise. The noise of the measurement from the $i^{th}$ to the $j^{th}$ agent is specified as $v_{ij}(t)$ which is assumed to be independent and identically distributed (i.i.d.) Gaussian random variables with zero mean and covariance $\mathrm{\Xi}_{ij}(t)\succ 0 $. Further the measurement and the noise sets of the $i^{th}$ agent are denoted by $Z_{i}(t)=\left[z_{i1}^{\mathrm{T}}(t)\cdots z_{iN}^{\mathrm{T}}(t) \right]^{\mathrm{T}}$ and $v_{i}(t) =\left[v_{i1}^{\mathrm{T}}(t) \cdots v_{iN}^{\mathrm{T}}(t) \right]^{\mathrm{T}}$, respectively. Over a finite time horizon $t=0, \cdots, T$, one can rewrite \eqref{eq:mas_dynamic} as a static form by stacking up the variables and matrices \citep{furieri2020first}: 
\begin{equation}\label{eq:state=dist+input}
\begin{split}
    x &= P_{11}w + P_{12}u\\
\end{split}
\end{equation}
where 
\begin{equation*}
    \begin{split}
    P_{11}&= (I-\mathrm{D}\bar{A})^{-1},\
    P_{12}= (I-\mathrm{D}\bar{A})^{-1}\mathrm{D}\bar{B}\\
        \bar{A} &= I_{T+1} \otimes \tilde{A},\ \ \ \ \ 
        \bar{B} = \left[\begin{matrix}
        I_{T}\otimes\tilde{B}\\
        0_{Nn\times NpT}
        \end{matrix}\right]\\  \mathrm{D}&=\left[\begin{matrix}
        0_{Nn\times NnT} &0_{Nn\times Nn}\\
        I_{NnT} &0_{NnT\times Nn}
        \end{matrix}\right]
    \end{split}
\end{equation*}
where $I_{T}$ and $0_{T}$ respectively denote the identity and zero matrices of size $T \times T$ , and $M_{i} = [0_{p} \cdots I_{p} \cdots 0_{p}]\in\mathbb{R}^{p\times Np}$ is the block matrix having $I_{p}$ in the $i^{th}$ block entry and filled with $0_{p}$ in other block entries. And $x = [x(0)^{\mathrm{T}} \dots x(T)^{\mathrm{T}}]^{\mathrm{T}}\in\mathbb{R}^{Nn(T+1)}$, and $u = \sum_{i}^{N}(I_{T}\otimes M_{i}^{\mathrm{T}})u_{i}\in\mathbb{R}^{NpT}$ are the stacked agents' states and their control inputs over the horizon, where $u_{i} = [u_{i}(0)^{\mathrm{T}} \dots u_{i}(T-1)^{\mathrm{T}}]^{\mathrm{T}}\in\mathbb{R}^{pT}, \forall i \in\mathcal{V}$. $w=[ x(0)^{\mathrm{T}} \ \tilde{w}(0)^{\mathrm{T}}\ \dots \tilde{w}(T-1)^{\mathrm{T}}]\in\mathbb{R}^{Nn(T+1)}$ is the vector containing initial agents' states and the additive noise.
Over the finite time horizon $T$, individual agents interact with their neighbors according to the control law $u_{i}$ embedded in each agent. Without loss of generality, $u_{i}$ can be designed by the following output feedback control law \citep{furieri2020first}: 
\begin{equation}\label{eq:measurement_input_equation}
    u_{i} = (I_{T}\otimes M_{i})\mathcal{F}Z_{i,(0:T-1)} =  (I_{T}\otimes M_{i})\mathcal{F}C(x+v_{i}), \ \ \forall i \in \mathcal{V}
\end{equation}
where $v_{i}= [v_{i}(0)^{\mathrm{T}} \dots v_{i}(T)^{\mathrm{T}}]^{\mathrm{T}}\in\mathbb{R}^{Nn(T+1)}, \forall i \in\mathcal{V}$, $Z_{i,(0:T-1)} = [Z_{i}(0)^{\mathrm{T}} \cdots Z_{i}(T-1)^{\mathrm{T}}]\in\mathbb{R}^{NnT}$, and $C =\left[I_{NnT} \ 0_{NnT\times Nn}\right]$. The crucial part is the design of the feedback gain, which is denoted by $\mathcal{F}\in\mathbb{F}$. Here, $\mathbb{F}\subset \mathbb{R}^{NpT \times NnT}$ is an invariant subspace that encodes network topological constraints for distributed MAS imposed by $\mathcal{A}$, as well as embeds causal feedback policies by forcing the future response entries to zeros.

\subsection{Optimal MAS Distributed Control Problem}
\begin{figure}[htpb]
\centering
\includegraphics[width=0.6\textwidth]{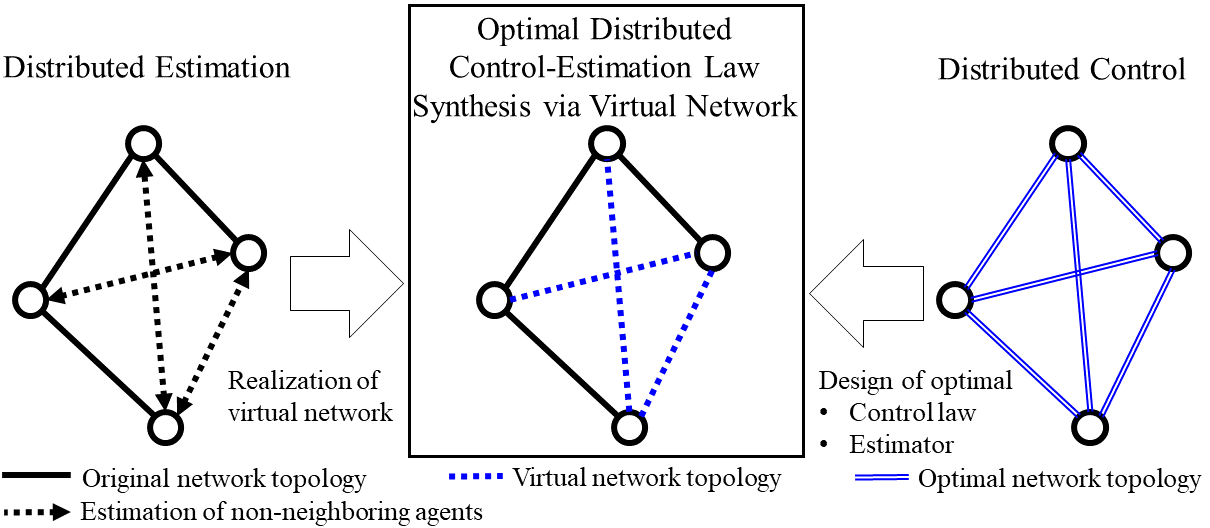}
\caption{Virtual interaction based distributed control-estimation synthesis.}\label{fig:finite_design}
\end{figure}

Given the equivalent static form of the stochastic linear MAS dynamics over the time horizon $T$ \eqref{eq:state=dist+input}, we seek to address the optimal distributed control problem.
 \begin{problem}\label{problem1}{\emph{Optimal distributed control law subject to structural constraint \citep{furieri2020first}:}} 
\begin{equation}\label{eq:cost_reformulation}
\begin{split}
\ \ \   & \min\limits_{\mathcal{F}\in\mathbb{F}} \  \mathbb{E}\left[x^{\mathrm{T}}\mathcal{Q}x+u^{\mathrm{T}}\mathcal{R}u\right] \\
 \mathrm{subje}  \mathrm{ct\ to} \ &\eqref{eq:state=dist+input}, \eqref{eq:measurement_input_equation},\ \ \ \ \   \forall i\in\mathcal{V}
\end{split}
\end{equation}
where $\mathcal{Q}\in\mathbb{R}^{Nn(T+1)\times Nn(T+1)}\succeq 0$, and  $\mathcal{R} \in\mathbb{R}^{NpT\times NpT}\succ 0$ are the associated weight matrices.
\end{problem}
Due to the structural constraints imposed on the control input space $\mathbb{F}$, Problem \ref{problem1} is a highly non-convex problem, which is indeed NP-hard and a formidable computational burden \citep{gupta2005sub}. To circumvent such a difficulty, we propose a concept of $virtual\ network\ topology$ that allows for the interactions between non-neighboring agents, i.e., $virtual\ interaction$ as depicted in Figure. \ref{fig:finite_design}. 

\begin{figure}[htpb]
\centering
\includegraphics[width=0.5\textwidth]{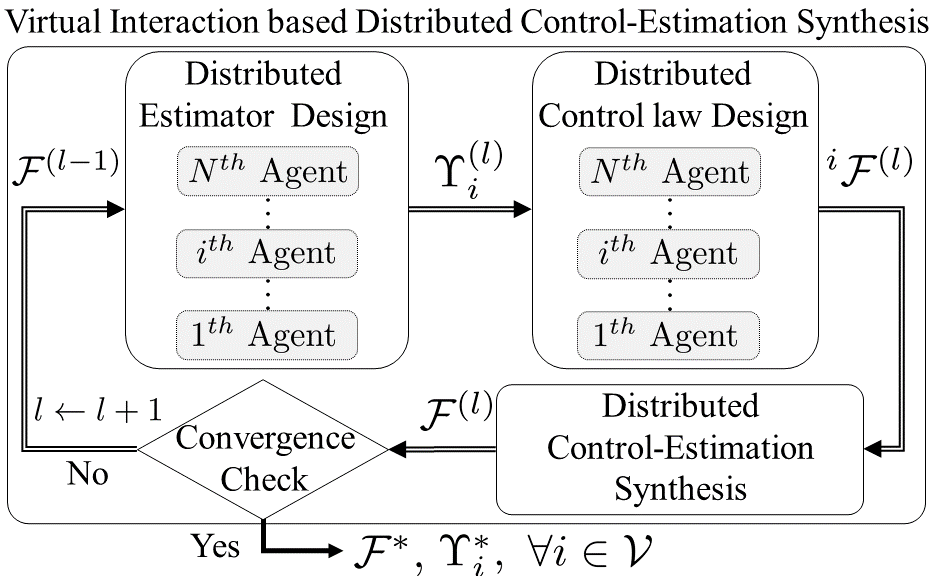}
\caption{Iterative optimization procedure for the optimal distributed control-estimation synthesis.}\label{fig:algorithm_diagram}
\end{figure}

Since the state information of the non-neighboring agent is not available, an appropriate estimator is required for each agent to obtain the estimates of non-neighboring agents' states. Using the Bayesian approach, Kalman-like filter is adopted for estimation as we consider a linear MAS along with the Gaussian uncertainties.
\\ 
\begin{definition} \label{def:estimate}
The state estimate and its covariance of the MAS using the $i^{th}$ agent's measurements are denoted by ${}^{i}\hat{x}(t):=\mathbb{E}\left[x(t)|Z_{i,(0:t)}\right]$ and  ${}^{i}\Sigma(t):=\mathbb{E}\left[\left(x(t)-{}^{i}\hat{x}(t)\right)\left(x(t)-{}^{i}\hat{x}(t)\right)^{\mathrm{T}}|Z_{i,(0:t)}\right], \forall j \in \mathcal{V}$, respectively \citep{kwon2018sensing}, where $\mathbb{E}[\bullet|\bullet]$ is the conditional expectation. 
\end{definition}
$ $  
The nominal recursive structure of Kalman-like filter is represented by: 
\begin{equation}\label{eq:KFstructure}
\begin{split}
    {}^{i}\hat{x}(t) & = {}^{i}\hat{x}^{-}(t)+L_{i}(t)H_{i}\left(Z_{i}(t)-{}^{i}\hat{x}^{-}(t)\right) 
\end{split}
\end{equation}
where ${}^{i}\hat{x}^{-}(t):=\mathbb{E}\left[x(t)|Z_{i,(0:t-1)}\right]$ denotes the predicted state estimate from the $i^{th}$ agent's perspective. $H_{i}\in\mathrm{R}^{n|\Omega_{i}| \times nN}$ only encodes the neighbor of the $i^{th}$ agent, that is, $H_{i} = [h_{1} \ h_{2}\ \cdots h_{|\Omega_{i}|}]^{\mathrm{T}} \otimes I_{n}$ where $h_{m} \in \mathbb{R}^{N}$, and $m={1,2,...,|\Omega_{i}|}$ are the nonzero column vectors of the matrix $diag(c_{i1},c_{i2}....,c_{iN})$. And, $L_{i}(t)\in \mathbb{R}^{nN \times n|\Omega_{i}|}$ represents the estimator gain at time step $t$ for estimating the states of the MAS from the $i^{th}$ agent's perspective. Once the entire MAS state estimate information becomes available for each agent, one can replace \eqref{eq:measurement_input_equation} with the estimation-based feedback control law. Accordingly, Problem \ref{problem1}, distributed control law subject to structural constraint, can be reformulated into the problem that simultaneously designs both distributed control and distributed estimator.
 \begin{problem}\label{problem2}{\emph{Optimal distributed control-estimation law with virtual interactions:}}
\begin{equation}\label{eq:problem2}
\begin{split}
\ \ \   & \min\limits_{\mathcal{F}\in\tilde{\mathbb{F}}, \Upsilon_{i},\forall i \in\mathcal{V}} \ J(\mathcal{F},\Upsilon_{1},\cdots,\Upsilon_{N}) \\
 \mathrm{subje}&  \mathrm{ct\ to} \ \eqref{eq:state_equation}, \ \mathrm{and}\\   
 & u_{i}=(I_{T}\otimes M_{i})\mathcal{F}C\ {}^{i}\hat{x},\ \ \ \ \  \forall i\in\mathcal{V}
\end{split}
\end{equation}
where ${}^{i}\hat{x}=[{}^{i}\hat{x}(0)^{\mathrm{T}} \cdots {}^{i}\hat{x}(T)^{\mathrm{T}}]^{\mathrm{T}}$, $J(\mathcal{F},\Upsilon_{1},\cdots,\Upsilon_{N}) :=\mathbb{E}\left[x^{\mathrm{T}}\mathcal{Q}x+u^{\mathrm{T}}\mathcal{R}u\right]$ and $\Upsilon_{i}:=\{L_{i}(t)| t=0,\cdots, T\}$ is the set of estimator gains over the time horizon $T$ for the $i^{th}$ agent.
\end{problem}
\begin{remark}\label{rm:subspaceFdefine}
It is worth noting that, compared to $\mathbb{F}$, $\tilde{\mathbb{F}}\subset \mathbb{R}^{NpT\times NnT}$ is a subspace that only encodes causal feedback policies, not restricted by any network topological constraint.
\end{remark}
Albeit Problem \ref{problem2} can successfully relax the structural constraint on the control law $\mathcal{F}$, it is not straightforward to solve as the control law and the state estimation errors mutually affect each other \citep{kwon2018sensing}. To resolve such complexity, we propose an iterative optimization procedure in a distributed fashion such that: i) divide the primal problem (Problem \ref{problem2}) into the set of sub-problems, each is viewed from an individual agent's perspective; ii) sequentially design the distributed estimation and control laws for each sub-problem; iii) mix the results from individual sub-problems to approximate the optimal solution to the primal problem. The overall schematic of the proposed distributed control-estimation synthesis is delineated in Figure. \ref{fig:algorithm_diagram}. 

For the $l^{th}$ iteration, the optimization procedure consists of the following sub-steps. Firstly, \emph{distributed estimator design} optimizes a set of estimator gains $\Upsilon_{i}^{(l)},\forall i\in\mathcal{V}$ based on the disturbance/noise model, the network topological constraint, and the suboptimal control law resulted from the previous iteration. Then, \emph{distributed control law design} computes a set of optimal control laws ${}^{i}\mathcal{F}^{(l)},\forall i\in\mathcal{V}$, each from the individual agents’ perspectives, based on the state estimation error information from the designed distributed estimator. Finally, \emph{distributed control-estimation synthesis} 
mixes the set of ${}^{i}\mathcal{F}^{(l)}, \forall i \in \mathcal{V}$ to construct the solution candidate, $\mathcal{F}^{(l)}$, for Problem \ref{problem2}. The constructed control law is evaluated to check the convergence and is used for the next iteration. The iteration terminates once the pre-defined stopping criteria are fulfilled, yielding the optimal control-estimation law, denoted by $\mathcal{F}^{*}$ and $\Upsilon_{i}^{\ast},\forall i\in\mathcal{V}$.
 
\section{Algorithm Development}\label{sc:algorithm}
This section details out the proposed synthesis procedure of the optimal distributed control-estimation law that can comply with an arbitrary network topology of the stochastic MAS.
\subsection{Distributed estimator design}\label{ssc:distest}
To begin with, the distributed estimation algorithm is optimized by means of estimator gains $\Upsilon_{i}^{(l)},\forall i\in\mathcal{V}$. As offline design phase, individual estimators can be designed based on the entire MAS model information along with the control law computed from the previous iteration, ($A$, $B$, and $\mathcal{F}^{(l-1)}$). For brevity, we use $\mathcal{F}$ to designate $\mathcal{F}^{(l-1)}$ in this subsection. Recalling (\ref{eq:KFstructure}), the distributed estimator embedded in the $i^{th}$ agent calculates ${}^{i}\hat{x}(t),\forall t\in\{0,\cdots,T\}$, whose performance can be measured by the estimation error.
\\
\begin{definition} \label{def:error}
${}^{i}e(t):=x(t)-{}^{i}\hat{x}(t)$ is the MAS state estimation error from the $i^{th}$ agent's perspective, and its covariance is ${}^{i}\Sigma(t)$ by Definition \ref{def:estimate}. Further, $e(t)=[{}^{1}e(t)^{\mathrm{T}}\cdots {}^{N}e(t)^{\mathrm{T}}]^{\mathrm{T}}\in\mathbb{R}^{NnN}$ stacks all the estimation errors from individual agents' estimator, and the corresponding covariance is denoted by $\Sigma(t):=\mathbb{E}[e(t)e(t)^{\mathrm{T}}]\in\mathbb{R}^{NnN\times NnN}$. Similarly, ${}^{i}e^{-}(t)$, ${}^{i}\Sigma^{-}(t)$, 
$e^{-}(t)$, $\Sigma^{-}(t)$ are defined in terms of the predicted state estimate ${}^{i}\hat{x}^{-}(t)$ \citep{kwon2018sensing}.
\end{definition}

\begin{assumption}\label{ass:initialestcov}
The initial conditions ${}^{i}\hat{x}(0),\forall i \in \mathcal{V}$, and $\Sigma(0)$ are given to individual agents in order to initiate each of their distributed estimators. 
\end{assumption}

Based on the prior knowledge, the estimation based control input of the $i^{th}$ agent at time step $t$ can be written by: 
\begin{equation}\label{eq:finite_input}
    u_{i}(t) = M_{i}\sum_{k=0}^{t}\mathcal{F}_{kt} \ {}^{i}\hat{x}(k)
\end{equation}
where $\mathcal{F}_{kt}\in\mathbb{R}^{pN\times nN}$ is the block matrix which spans from $(knN)^{th}$ to $(knN+nN-1)^{th}$ columns and from $(kpN)^{th}$ to $(kpN+pN-1)^{th}$ rows of the control law matrix $\mathcal{F}$. With \eqref{eq:finite_input}, the entire MAS dynamics \eqref{eq:mas_dynamic} can be expressed by:
\begin{equation}\label{eq:total_dynamic3}
\begin{split}
      x(t+1) & = \tilde{A}x(t)+ \tilde{B}\mathcal{F}_{tt}x(t) 
     +\sum_{k=0}^{t-1}\tilde{B}\mathcal{F}_{kt}x(k)  - \sum_{k=0}^{t}\bar{B}\tilde{M}\tilde{\mathcal{F}}_{kt}e(k)+\tilde{w}(t)
\end{split}
\end{equation}
where $\bar{B} = 1_{N}^{\mathrm{T}} \otimes \tilde{B}$, $\tilde{M} = blkdg(M_{1}^{\mathrm{T}}M_{1},\dots,M_{N}^{\mathrm{T}}M_{N})\in\mathbb{R}^{NpN\times NpN}$, and $\tilde{\mathcal{F}}_{kt} = I_{N} \otimes \mathcal{F}_{kt}$. $blkdg(\bullet)$ denotes a block-diagonal matrix with block matrices $\bullet$, and the vector $1_{N}\in\mathbb{R}^{N}$ indicates every element equals to $1$. Then the predicted state estimate of the entire MAS from the $i^{th}$ agent's perspective is given by: 
\begin{equation}\label{eq:predicted_state}
    \begin{split}
        {}^{i}\hat{x}^{-}(t+1) = \tilde{A}\ {}^{i}\hat{x}(t)+ \tilde{B}\mathcal{F}_{tt}{}^{i}\hat{x}(t) 
     +\sum_{k=0}^{t-1}\tilde{B}\mathcal{F}_{kt}{}^{i}\hat{x}(k)
    \end{split}
\end{equation}
Subtracting \eqref{eq:predicted_state} from  \eqref{eq:total_dynamic3}, and concatenating the results for all agents in $\mathcal{V}$ gives: 
\begin{equation}\label{eq:mas_p_err}
\begin{split}
e^{-}(t+1) &= \Lambda_{t}e(t)+\sum_{k=0}^{t-1}\Psi_{kt}e(k)+1_{N}\otimes\tilde{w}(t)\\
\mathrm{where}\ \Lambda_{t} &= I_{N}\otimes (\tilde{A}+\tilde{B}\mathcal{F}_{tt})-1_{N} \otimes \bar{B}\tilde{M}\tilde{\mathcal{F}}_{tt},\ \  
      \Psi_{kt} = I_{N}\otimes \tilde{B}\mathcal{F}_{kt}- 1_{N} \otimes \bar{B}\tilde{M}\tilde{\mathcal{F}}_{kt}
\end{split}
\end{equation}
Correspondingly, $\Sigma^{-}(t+1)$ is represented by:
\begin{equation}\label{eq:PredictionCov}
    \begin{split}
        \Sigma^{-}(t+1) & = \Lambda_{t}\Sigma(t)\Lambda_{t}^{\mathrm{T}}+ \Sigma_{\tilde{w}}(t)
         + \sum_{q=0}^{t-1}\Lambda_{t}\mathbb{E}[e(t)e(q)^{\mathrm{T}}]\Psi_{qt}^{\mathrm{T}} +\sum_{p=0}^{t-1}\Psi_{pt}\mathbb{E}[e(p)e(t)^{\mathrm{T}}]\Lambda_{t}^{\mathrm{T}} 
          +\sum_{p=0}^{t-1}\sum_{q=0}^{t-1}\Psi_{pt}\mathbb{E}[e(p)e(q)^{\mathrm{T}}]\Psi_{qt}^{\mathrm{T}}
    \end{split}
\end{equation}
where $\Sigma_{\tilde{w}}(t)=(1_{N}1_{N}^{\mathrm{T}})\otimes blkdg(\Theta_{1}(t),\dots,\Theta_{N}(t))$. Note that the summation terms in the RHS of \eqref{eq:PredictionCov} (e.g., $\mathbb{E}[e(p)e(q)^{\mathrm{T}}], q\neq p$) imply the correlations of state estimates over time induced by the control law \eqref{eq:finite_input}. Now, the predicted error, (\ref{eq:KFstructure}) can be rewritten by:
\begin{equation}\label{eq:meanupdate}
\begin{split}
    {}^{i}\hat{x}(t+1) =&{}^{i}\hat{x}^{-}(t+1) 
    +L_{i}(t+1)H_{i}({}^{i}e^{-}(t+1)+v_{i}(t+1))
\end{split}
\end{equation}
Like the Kalman gain, $L_{i}(t+1)$ can be computed in a way minimizing the mean-squared error of the state estimate, i.e., $\mathbb{E}\left[\left\|{}^{i}e(t+1)\right\|^{2}\right]$. This is in fact equivalent to minimizing the trace of the posterior covariance matrices, i.e., $\mathrm{Tr}\left({}^{i}\Sigma(t+1)\right), \forall i \in \mathcal{V}$. By the definition of ${}^{i}\Sigma(t+1)$, we have:
\begin{equation}\label{eq:cov_update}   
    \begin{split}
        {}^{i}\Sigma&(t+1):=\mathbb{E}[{}^{i}e(t+1){}^{i}e(t+1)^{\mathrm{T}}|Z_{i,(0:t+1)}]
        \\ 
        =&\left(I_{nN}-L_{i}(t+1)H_{i}\right){}^{i}\Sigma^{-}(t+1)(I_{nN}-L_{i}(t+1)H_{i})^{\mathrm{T}}  +L_{i}(t+1)H_{i}{}^{i}\Xi(t+1)(L_{i}(t+1)H_{i})^{\mathrm{T}}
    \end{split}
\end{equation}
       where     
     \begin{equation}\label{eq:optestgain}
    \begin{split}   
          L_{i}(k+1) &={}^{i}\Sigma^{-}(t+1)H_{i}^{\mathrm{T}}(S_{i}(t+1))^{-1} \\
 S_{i}(t+1)&=H_{i}({}^{i}\Sigma^{-}(t+1)+{}^{i}\mathrm{\Xi}(x+1) )H_{i}^{\mathrm{T}} \\
 {}^{i}\mathrm{\Xi}(t+1)&=blkdg(\mathrm{\Xi}_{i1}(t+1), \ \mathrm{\Xi}_{i2}(t+1), \dots \mathrm{\Xi}_{iN}(t+1))
    \end{split}
\end{equation}
Correspondingly, $\Sigma(t+1)$ can be updated by:
\begin{equation}\label{eq:posteriorCov}   
\begin{split}
 \Sigma(t+1)&=(I-\tilde{L}(t+1)) \Sigma^{-}(t+1)(I-\tilde{L}(t+1))^{\mathrm{T}}  + \tilde{L}(t+1)\Sigma_{\Xi}(t+1)\tilde{L}(t+1)^{\mathrm{T}}
\end{split}    
\end{equation}
where $\Sigma_{\Xi}(t+1)=blkdg({}^{1}\mathrm{\Xi}(t+1),\dots, {}^{N}\mathrm{\Xi}(t+1))$, and $\tilde{L}(t+1)= blkdg(L_{1}(t+1)H_{1}, \dots, L_{N}(t+1)H_{N})$. Note that, the covariance between the state estimation errors at current and past steps, i.e.,  $\mathbb{E}[e(t+1)e(s)^{\mathrm{T}}]$ and $\mathbb{E}[e(s)e(t+1)^{\mathrm{T}}],\forall s<t$ need to be updated using the computed $\tilde{L}(t+1)$. The cross-covariance between the $i^{th}$ and the $j^{th}$ agents' state estimates ${}^{ij}\Sigma(t+1):=\mathbb{E}[{}^{i}e(t+1){}^{j}e(t+1)^{\mathrm{T}}]\in\mathbb{R}^{Nn\times Nn}$ is at the off-diagonal block entry, while ${}^{i}\Sigma(t+1)\in\mathbb{R}^{Nn\times Nn}$ is at the diagonal block entry of $\Sigma(t+1)\in\mathbb{R}^{NnN\times NnN}$. The detailed expansions of $\Sigma(t)$ is as follows:
\begin{equation*}
    \begin{split}
        \Sigma(t) &= \left[
\begin{matrix}
{}^{1}\Sigma(t)   &{}^{12}\Sigma(t)  &\cdots &{}^{1N}\Sigma(t)\\ 
{}^{21}\Sigma(t) &{}^{2}\Sigma(t) &\cdots &{}^{2N}\Sigma(t)\\
\vdots                                   &\vdots              &\ddots   &\vdots\\ 
{}^{N1}\Sigma(t)        &{}^{N2}\Sigma(t)  &\cdots  &{}^{N}\Sigma(t)
\end{matrix}\right]
    \end{split}
\end{equation*}
Based on the computed $\Upsilon_{i}$ from \eqref{eq:optestgain}, each agent can update ${}^{i}\hat{x}(t)$, ${}^{i}\Sigma(t)$ and $\Sigma(t)$, respectively using \eqref{eq:meanupdate}, \eqref{eq:cov_update}, and \eqref{eq:posteriorCov}. This completes the implementation of the distributed estimation algorithm. 
\\
\begin{remark}\label{rm:covissame}
It is noted that $\Sigma(t)$ computed by each agent is irrespective of agent's perspective since the same initial condition $\Sigma(0)$ is given to each agent by Assumption \ref{ass:initialestcov}.
\end{remark}

\subsection{Distributed control law design}\label{ssc:distctrl}
In this section, the computationally tractable formulation of the optimal distributed control law is derived from the individual agents' perspectives. The main idea starts with relaxing the structural constraints on $\mathcal{F}$ by applying the estimator \eqref{eq:meanupdate} to each agent.

\begin{definition} \label{def:time_series_error}
Let ${}^{i}e:= x-{}^{i}\hat{x}$ stacks up the time series of the estimation errors from the $i^{th}$ agent's perspective, over the time horizon $T$. Given $\mathcal{F}$ and $\Upsilon_{i}, \forall i\in\mathcal{V}$, one can construct the estimation error covariance over the time horizon $T$, $\Sigma_{i}:=\mathbb{E}[{}^{i}e{}^{i}e^{\mathrm{T}}],\forall i\in\mathcal{V}$, as well as the cross-covariance between two different agents $\Sigma_{ij}:=\mathbb{E}[{}^{i}e\ {}^{j}e^{\mathrm{T}}], \forall i \neq j \in\mathcal{V}$.
\end{definition}

In terms of the time series of the estimation errors, the state estimation based control law over the time horizon can be expressed by:
\begin{equation}\label{eq:u_x_e}
    u = \sum_{i}^{N} \mathcal{M}_{i}\mathcal{F}C\ {}^{i}\hat{x} = \mathcal{F}Cx-\sum_{i}^{N} \mathcal{M}_{i}\mathcal{F}C\,{}^{i}e
\end{equation}
where $\mathcal{M}_{i} = I_{T}\otimes (M_{i}^{\mathrm{T}}M_{i}),\forall i\in\mathcal{V}$. Plugging \eqref{eq:u_x_e} into \eqref{eq:state=dist+input} yields the objective cost in \eqref{eq:problem2} as follows:
\begin{equation}\label{eq:distest_cost}
    \begin{split}  
J(\mathcal{F},\Upsilon_{1},\cdots,\Upsilon_{N})  =&  \Vert  \mathcal{Q}^{\frac{1}{2}}(I-P_{12}\mathcal{F}C)^{-1}P_{11}\Sigma_{w}^{\frac{1}{2}}\Vert^2_{{}_{F}}
 +\Vert \mathcal{R}^{\frac{1}{2}}(I-\mathcal{F}CP_{12})^{-1}\mathcal{F} C P_{11}\Sigma_{w}^{\frac{1}{2}}\Vert^2_{{}_{F}}\\
    & +\sum_{i,j}
    \Vert \mathcal{Q}^{\frac{1}{2}}P_{12} (I-\mathcal{F}CP_{12})^{-1}(\mathcal{M}_{i}\mathcal{F}C\Sigma_{ij}C^{\mathrm{T}}\mathcal{F}^{\mathrm{T}}\mathcal{M}_{j}^{\mathrm{T}})^{\frac{1}{2}}\Vert^2_{{}_{F}} \\
     &+\sum_{i,j}\Vert \mathcal{R}^{\frac{1}{2}}(I-\mathcal{F}CP_{12})^{{}^{-1}}(\mathcal{M}_{i}\mathcal{F}C\Sigma_{ij}C^{\mathrm{T}}\mathcal{F}^{\mathrm{T}}\mathcal{M}_{j}^{\mathrm{T}})^{\frac{1}{2}}\Vert^2_{{}_{F}}  \\
    & +\Vert \mathcal{Q}^{\frac{1}{2}}(I-P_{12}\mathcal{F}C)^{-1}P_{11}\mu_{w}
    \Vert^{2}_{2} +\Vert \mathcal{R}^{\frac{1}{2}}(I-\mathcal{F}CP_{12})^{-1}\mathcal{F} CP_{11}\mu_{w}\Vert^{2}_{2}
    \end{split}
\end{equation}
where $\Vert \cdot \Vert^{2}_{2}$ and $\Vert \cdot \Vert^{2}_{{}_{F}} $ denote Euclidean norm and the Frobenius norm, respectively; and $\mu_{w}:=\mathbb{E}[w]\in\mathbb{R}^{Nn(T+1)}$, $\Sigma_{w}:=\mathbb{E}[(w-\mu_{w})(w-\mu_{w})^{\mathrm{T}}]\in\mathbb{R}^{Nn(T+1)\times Nn(T+1)}$.

Apparently, the objective cost \eqref{eq:distest_cost} has high-dimensional, highly coupled optimization variables, i.e., $\mathcal{F}$, which is our main interest, and $\Sigma_{ij}, \forall i,j\in\mathcal{V}$, which are the implicit functions of both $\mathcal{F}$ and $\Upsilon_{i}, \forall i\in\mathcal{V}$. The proposed iterative optimization procedure alleviates these coupling complexities in two aspects. First, akin to the alternating direction method of multipliers (ADMM) technique \citep{lin2013design}, we set $\Sigma_{ij}, \forall i,j\in\mathcal{V}$ constant while optimizing $\mathcal{F}$ at the $l^{th}$ iteration, thereby treating $J$ as the function of $\mathcal{F}$ only. Note that $\Upsilon_{i}, \forall i\in\mathcal{V}$ is designed over the constant $\mathcal{F}$ in the \emph{distributed estimator design} phase of the next iteration. Second, we interpret the global objective cost from the individual agent's viewpoint, and translate the primal problem (Problem \ref{problem2}) into the agent-wise objective cost. The objective cost, locally seen by the $i^{th}$ agent's viewpoint at the $l^{th}$ iteration, is denoted by ${}^{i}J^{(l)}$ which can be constructed using the estimated MAS input ${}^{i}u^{(l)} = {}^{i}\mathcal{F}^{(l)}C(x-{}^{i}e)$ instead of \eqref{eq:u_x_e}. ${}^{i}\mathcal{F}^{(l)}$ is constructed from the $i^{th}$ agent's perspective by optimizing the agent-wise objective cost, ${}^{i}J^{(l)}$. Then, the resulting agent-wise optimization problem is represented as follows: 
\begin{problem}\label{problem3}{\emph{Optimal distributed control law from agent-wise viewpoint:}} 
\begin{equation}\label{eq:cost_reformulation}
\begin{split}
\ \ \   & \min\limits_{{}^{i}\mathcal{F}^{(l)}\in\mathbb{\tilde{F}}} \  {}^{i}J^{(l)}({}^{i}\mathcal{F}^{(l)}) 
\end{split}
\end{equation}
where
\begin{equation}\label{eq:localcost}
    \begin{split} 
 {}^{i}J^{(l)}({}^{i}\mathcal{F}^{(l)}) =&  \Vert  \mathcal{Q}^{\frac{1}{2}}(I-P_{12}\,{}^{i}\mathcal{F}^{(l)}C)^{^{-1}}P_{11}{}^{i}\Sigma_{w}^{\frac{1}{2}}\Vert^2_{{}_{F}} +\Vert \mathcal{R}^{\frac{1}{2}}(I-{}^{i}\mathcal{F}^{(l)}CP_{12})^{^{-1}}\,{}^{i}\mathcal{F}^{(l)} C P_{11}{}^{i}\Sigma_{w}^{\frac{1}{2}}\Vert^2_{{}_{F}}\\
&   +\Vert \mathcal{Q}^{\frac{1}{2}}(I-P_{12}\,{}^{i}\mathcal{F}^{(l)}C)^{^{-1}}P_{12}\,{}^{i}\mathcal{F}^{(l)}C\Sigma_{i}^{(l)\frac{1}{2}}\Vert^2_{{}_{F}}  + \Vert \mathcal{R}^{\frac{1}{2}}(I-{}^{i}\mathcal{F}^{(l)}CP_{12})^{^{-1}}\,{}^{i}\mathcal{F}^{(l)}C\Sigma_{i}^{(l)\frac{1}{2}}\Vert^2_{{}_{F}}\\
    & +\Vert \mathcal{Q}^{\frac{1}{2}}(I-P_{12}\,{}^{i}\mathcal{F}^{(l)}C)^{^{-1}}P_{11}{}^{i}\mu_{w}
    \Vert^{2}_{2}   +\Vert \mathcal{R}^{\frac{1}{2}}(I-{}^{i}\mathcal{F}^{(l)}CP_{12})^{^{-1}}\,{}^{i}\mathcal{F}^{(l)} CP_{11}{}^{i}\mu_{w}\Vert^{2}_{2}
    \end{split}
\end{equation}
where ${}^{i}\mu_{w}=\mathbb{E}[w|Z_{i,(0:T)}]$, ${}^{i}\Sigma_{w}:=\mathbb{E}[(w-{}^{i}\mu_{w})(w-{}^{i}\mu_{w})^{\mathrm{T}}|Z_{i,(0:T)}],\forall i\in\mathcal{V}$. Note that  $\Sigma_{i}^{(l)}\in\mathbb{R}^{Nn(T+1)\times Nn(T+1)}$ is computed by Definition \ref{def:time_series_error} at the $l^{th}$ iteration.
\end{problem} 

$ $
\begin{definition}
A subspace $\tilde{\mathbb{F}}$ is quadratic invariance (QI) with respect to $CP_{12}$ if and only if ${}^{i}\mathcal{F}^{(l)}CP_{12}{}^{i}\mathcal{F}^{(l)}\in\tilde{\mathbb{F}}$. And it is trivial to show that $\tilde{\mathbb{F}}$ is QI with respect to $CP_{12}$\citep{lessard2011quadratic}. 
\end{definition}
It is well-known fact that QI is a sufficient and necessary condition for the exact convex reformulation \citep{lessard2011quadratic}. That is, one can apply an equivalent disturbance-feedback policy to make \eqref{eq:localcost} a convex form, similar to \citep{furieri2020first}. 
\\

\begin{definition}\label{lm:non_lin_mapping}
Let us introduce the nonlinear mapping as:
\begin{equation}\label{eq:non_lin_mapping}
h(\Phi) = (I+\Phi CP_{12})^{-1}\Phi, \ \ h: \mathbb{R}^{NpT\times NnT} \mapsto \mathbb{R}^{NpT\times NnT}
\end{equation}
and define the cost function $\tilde{J}:\mathbb{R}^{NpT\times NnT}\mapsto \mathbb{R}$ in terms of the design parameter ${}^{i}\Phi^{(l)}$\citep{furieri2020first}.
\begin{equation}\label{eq:local_cost_cvx}
    \begin{split} 
{}^{i}\tilde{J}^{(l)}({}^{i}\Phi^{(l)}) &=  \Vert  \mathcal{Q}^{\frac{1}{2}}(I+P_{12}{}^{i}\Phi^{(l)} C)P_{11}{}^{i}\Sigma_{w}^{\frac{1}{2}}\Vert^2_{{}_{F}} +\Vert \mathcal{R}^{\frac{1}{2}}{}^{i}\Phi^{(l)} CP_{11}{}^{i}\Sigma_{w}^{\frac{1}{2}}\Vert^2_{{}_{F}} +\Vert \mathcal{Q}^{\frac{1}{2}}P_{12}{}^{i}\Phi^{(l)}\Sigma_{i}^{(l)\frac{1}{2}}\Vert^2_{{}_{F}} \\
& \ \ \ + \Vert \mathcal{R}^{\frac{1}{2}}{}^{i}\Phi^{(l)}\Sigma_{i}^{(l)\frac{1}{2}}\Vert^2_{{}_{F}}+\Vert \mathcal{R}^{\frac{1}{2}}{}^{i}\Phi^{(l)} CP_{11}{}^{i}\mu_{w}\Vert^{2}_{2}  +\Vert \mathcal{Q}^{\frac{1}{2}}(I+P_{12}{}^{i}\Phi^{(l)} C)P_{11}{}^{i}\mu_{w} \Vert^{2}_{2}
    \end{split}
\end{equation}

\end{definition}
By Theorem $1$ in \citep{furieri2020first}, the following convex optimization problem is equivalent to Problem \ref{problem3}.  
\\ 

\begin{problem}\label{problem4}{\emph{Equivalent convex problem to optimal distributed control from agent-wise viewpoint:}} 
\begin{equation}\label{eq:apprxconvexprob}
    \min\limits_{{}^{i}\Phi^{(l)}\in h^{-1}(\tilde{\mathbb{F}})} {}^{i}\tilde{J}^{(l)}({}^{i}\Phi^{(l)}) 
\end{equation}
\end{problem}
By solving \eqref{eq:apprxconvexprob} using convex programming, one can find the optimal ${}^{i}\Phi^{(l)}$, and the corresponding ${}^{i}\mathcal{F}^{(l)}$ from the inverse mapping $h^{-1}$ of \eqref{eq:non_lin_mapping}. The same optimization routines (Problem \ref{problem3} and \ref{problem4}) based on the locally seen cost from the other agents' perspectives are processed to get the optimal control laws ${}^{i}\mathcal{F}^{(l)},\forall i\in\mathcal{V}$ at the $l^{th}$ iteration step. 

\subsection{Distributed control-estimation synthesis}\label{ssc:subctrl_synthesis}
The set of optimal control laws from individual agents' viewpoint, ${}^{i}\mathcal{F}^{(l)},\forall i\in\mathcal{V}$, is mixed to approximate the solution to Problem \ref{problem2} by the agent-wise mixing strategy proposed as follows: 
\begin{equation}\label{eq:gainmerge}
\mathcal{F}^{(l)} = \sum_{i}^{N}\mathcal{M}_{i}\,{}^{i}\mathcal{F}^{(l)}
\end{equation} 
The basic intuition of the proposed strategy is to exhibit the control law for the $i^{th}$ agent using the one computed from the sup-optimization problem (Problem \ref{problem3}) from the $i^{th}$ agent's perspective. Accordingly, the proposed mixing strategy allows for individual agents to retain distributed controllers to be executed, retaining each of their sub-optimal solutions without interfering with each other.

\subsection{Convergence check}\label{ssc:eval}
The last step of the iteration loop evaluates the designed distributed control law \eqref{eq:gainmerge}, together with the estimator \eqref{eq:KFstructure}. First, let $S$ be a set which stores the designed control law, $\mathcal{F}^{(l)}$, and the estimator gains, $\Upsilon_{i}^{(l)},\ \forall i\in\mathcal{V}$, from each iteration step as follows:
\begin{equation}\label{eq:finite_cost_set}
S:=\left\{ s^{(l)} \bigg| s^{(l)}=\left(\mathcal{F}^{(l)},\Upsilon_{1}^{(l)},\cdots,\Upsilon_{N}^{(l)} \right),\ l \in \mathbb{N}\right\}
\end{equation}
The iteration terminates if: i) the total iteration counts
the threshold number $N_{max}$; or ii) the consecutive iteration is converged with respect to the following stopping condition:
\begin{equation}\label{eq:finite_stop}
    \triangle J(l,l-1)\le \epsilon_{stop} 
\end{equation}
where $\triangle J(l,l-1):= \lvert J(\mathcal{F}^{(l)},\Upsilon_{1}^{(l)},\cdots, \Upsilon_{N}^{(l)})-J(\mathcal{F}^{(l-1)},\Upsilon_{1}^{(l-1)},\cdots, \Upsilon_{N}^{(l-1)})\lvert$ and $\epsilon_{stop}$ is the threshold magnitude for the convergence. The objective cost of the corresponding control law $J(\mathcal{F}^{(l)},\Upsilon_{1}^{(l)},\cdots, \Upsilon_{N}^{(l)})$ is computed by plugging the designed control law, $\mathcal{F}^{(l)}$, and the set of distributed estimator gains $\Upsilon_{i}^{(l)},\forall i\in\mathcal{V}$ into \eqref{eq:distest_cost}. The final output of the control-estimation synthesis is given by:
\begin{equation}\label{eq:opt_sol}
\begin{split}
    \mathcal{F}^{\ast} =& \mathcal{F}^{(l)},\  \Upsilon_{i}^{\ast}=\Upsilon_{i}^{(l)},\ \forall i\in\mathcal{V}\\
    \mathrm{where}  \ l =& \displaystyle \argmin_{ \forall l \in |S|} J(\mathcal{F}^{(l)},\Upsilon_{1}^{(l)},\cdots, \Upsilon_{N}^{(l)}) 
\end{split}
\end{equation}
The overall recursive structure of the control-estimation synthesis procedure is summarized in Algorithm \ref{Algorithm_finite}. 

\begin{algorithm}[ht]
\textbf{Initialize} the MAS dynamics information $A$, $B$, $\mathcal{L}$, $\Sigma(0)$, $\mathcal{F}^{(0)}$, $\epsilon_{stop}$, $N_{max}$  and the cost metrics $\mathcal{Q}$, $\mathcal{R}$. \\
\textbf{for} $l=1,2,\cdots N_{max}$ do

\begin{enumerate}[a)]
\item{Distributed estimator design\\
\textbf{for} t=0 to the termination time $T$\\
\begin{enumerate}[1)]
\item{Update $\Sigma(t+1)$ using $\mathcal{F}^{(l-1)}$, \eqref{eq:PredictionCov}, \eqref{eq:optestgain}, and \eqref{eq:posteriorCov}}
\end{enumerate}}   
\textbf{end for}, Output $\longrightarrow$ $\Upsilon_{i}^{(l)}\ \mathrm{and}\ \Sigma_{i}^{(l)},\forall i\in\mathcal{V}$
\item{Distributed control law design\\
\textbf{for} i=1 to the number of total agent, $N$\\
\begin{enumerate}[2)]
\item{Solve \eqref{eq:apprxconvexprob}, and compute ${}^{i}\mathcal{F}^{(l)}$}
\end{enumerate}}
\textbf{end for}, Output $\longrightarrow$ ${}^{i}\mathcal{F}^{(l)},\forall i\in\mathcal{V}$

\item{Distributed control-estimation synthesis}
\begin{enumerate}[3)]
\item{Synthesize the control law $\mathcal{F}^{(l)}$ using \eqref{eq:gainmerge}}
\end{enumerate}

\item{Convergence check}

\begin{enumerate}[4)]
\item{Store $\mathcal{F}^{(l)}$, and $\Upsilon_{i}^{(l)},\forall i\in\mathcal{V}$ in the set $S$}
\end{enumerate}

\begin{enumerate}[5)]
\item{If \eqref{eq:finite_stop} is satisfied or $l>N_{max}$, then terminates.}
\end{enumerate}
\end{enumerate}
\textbf{end for}, \ \textbf{Output} $\Longrightarrow  \mathcal{F}^{\ast}$, and $\Upsilon_{i}^{\ast},\ \forall i\in\mathcal{V}$ 
\caption{{\bf Virtual interaction based distributed control-estimation synthesis.} \label{Algorithm_finite}}
\end{algorithm}

It is noted that the algorithm \ref{Algorithm_finite} is executed in the offline design phase. Once the distributed control law $\mathcal{F}^{\ast}$ and the corresponding estimator gains $\Upsilon_{i}^{\ast}$ for each agent are designed, each agent is deployed into the distributed online operation using its own prior knowledge. It is worth noting that the majority of the heavy computations occur at the offline design phase. Therefore, when it comes to the online operation, it is not burdensome to the limited on-board resources of each agent. Indeed, the computational complexity of the online operation for the proposed algorithm is scaled by the number of agents, i.e., $\mathcal{O}(N)$.

\section{Stability Analysis}\label{sc:analysis}
In this session, the stability analysis of the proposed distributed estimation algorithm in section \ref{ssc:distest} is presented. To begin with, 
let us consider the control law as static memoryless feedback gain $F$ as follows: 
\begin{equation}\label{eq:infinite_input_protocol}
u_{i}(t) = M_{i}F\, {}^{i}\hat{x}(t)
\end{equation}
where $M_{i}=[0_{p} \cdots I_{p} \cdots 0_{p}]\in\mathbb{R}^{p\times Np}$ is the block matrix having $I_{p}$ in the $i^{th}$ block entry and filled with $0_{p}$ in other entries. $F$ has structural constraints subject to the network topology of MAS specified by the Laplacian $\mathcal{L}$. Note that the estimation stability is unrelated to the design of $F$ as will be discussed below, and thus readily applicable to memory based feedback control law as in \eqref{eq:finite_input}. Corresponding to \eqref{eq:infinite_input_protocol}, the predicted state estimation errors of the $i^{th}$ agent can be written as follows:  
\begin{equation}\label{eq:agent_p_err_inf}
        {}^{i}e^{-}(t+1)  = D_{1}{}^{i}e(t) + D_{12}e(t)+\tilde{w}(t)
\end{equation}
where
\begin{equation*}
\begin{split}
         D_{1}  &= \tilde{A}+\tilde{B}F\\
         D_{12} &= -(1_{N}^{\mathrm{T}}\otimes \tilde{B})blkdg(M_{1}^{\mathrm{T}}M_{1},\dots,M_{N}^{\mathrm{T}}M_{N})(I_{N}\otimes F)\\
         e(t)&= [{}^{1}e(t)^{\mathrm{T}}\cdots {}^{N}e(t)^{\mathrm{T}}]^{\mathrm{T}}
\end{split}
\end{equation*}
It is noted from the above equation \eqref{eq:agent_p_err_inf} that the estimation error of the individual agent ${}^{i}e$ is coupled with the augmented estimation error of the entire MAS, $e$. Then, the following two lemmas are required for proving the stability of the proposed distributed estimation algorithm.

\begin{lemma}\label{lm:obsv_lm}
Estimation error covariance from the $i^{th}$ agent, ${}^{i}\Sigma(t)$ is positive definite and bounded for all $t$ if the following system is observable \citep{kwon2018sensing}:
\begin{equation}\label{eq:obsv_dynm}
    \begin{split}
        x(t+1) &= \mathcal{L}x(t)\\
        Z_{i}(t) &= C_{i}x(t)
    \end{split}
\end{equation}
\end{lemma}
\noindent where $C_{i}= [h_{1} \ h_{2}\ \cdots h_{|\Omega_{i}|}]^{\mathrm{T}} \otimes I_{n}\in\mathbb{R}^{n|\Omega_{i}| \times nN}$ is a observer matrix which gathers the measurements from the $i^{th}$ agent's perspective, i.e., those that are neighboring agents of the $i^{th}$ agent. $h_{q} \in \mathbb{R}^{N}, q={1,2,...,|\Omega_{i}|}$ are the nonzero column vectors of the matrix $diag(c_{i1},c_{i2}....,c_{iN})$. It is noted that the value of $c_{ij}$ can be decided by the graph $\mathcal{G}=(\mathcal{V},\mathcal{E})$ of the given network where $(i,j)\in \mathcal{E}$ indicates the availability of measurement of the $j^{th}$ agent's state from the $i^{th}$ agent, i.e., $c_{ij}=1$ and $(i,j)\notin \mathcal{E}$ means $c_{ij}=0$. 
\\ \\ 
\emph{Proof}.
The proof is referred to in the author's previous work \citep{kwon2018sensing}. \hfill\(\blacksquare\)

To analyze the estimation stability of ${}^{i}e$, we first introduce $H_{i}(t) \in \mathbb{R}^{nN^{2}\times nN}$ as a affine mapping matrix and $\alpha_{i}(t)\in\mathbb{R}^{nN^{2}}$ as a lumped noise as follows:
\begin{equation}
    \begin{split}
        H_{i}(t+1) &= \tilde{F}(t+1)H_{i}(t)({}^{i}D_{22}(t+1){}^{i}D_{2}(t))^{-1}   \\
        \alpha_{i}(t+1) &= (\tilde{F}(t+1)-H_{i}(t+1){}^{i}D_{22}(t+1)D_{12})\alpha_{i}(t)+\gamma(t+1)-H_{i}(t+1)\zeta_{i}(t+1)\\
        {}^{i}D_{2}(t) &= D_{1}+D_{12}H_{i}(t)\\
        {}^{i}D_{22}(t+1)&= I_{nN}-L_{i}(t+1)C_{i}\\
        \zeta_{i}(t+1) &= {}^{i}D_{22}(t+1)\tilde{w}(t)-L_{i}(t+1)C_{i}v_{i}(t+1)\\
        \tilde{F}(t+1) &= blkdg(
        {}^{1}D_{22}(t+1){}^{1}D_{2}(t),\  \dots,\ {}^{N}D_{22}(t+1){}^{N}D_{2}(t))\\
        \gamma(t+1)&= 
        \begin{bmatrix}
{}^{1}D_{22}(t+1)D_{12}\alpha_{1}(t)+\zeta_{1}(t+1)\\
\vdots\\
{}^{N}D_{22}(t+1)D_{12}\alpha_{N}(t)+\zeta_{N}(t+1)
\end{bmatrix}
    \end{split}
\end{equation}

\noindent It is noted that it is trivial to compute initial affine transformation matrix $H_{i}(0), \forall i \in \mathcal{V}$ which satisfies $\Sigma(0) = H_{i}{}^{i}\Sigma(0)H_{i}^{\mathrm{T}}$ under the Assumptions \ref{ass:initialestcov}. On the other hand, $\alpha_{i}$ follows the Gaussian distribution $\alpha_{i}(t) \sim \mathcal{N}_{nNN}(0,\eta_{i}(t))$, where $\eta_{i}(t)=\mathbb{E}[\alpha_{i}(t)\alpha_{i}^{\mathrm{T}}(t)]^{\mathrm{T}}$ with initial value as $\eta_{i}(0)=0_{nNN}$. Then, we can show that the augmented estimation error can be mapped to the estimation error from the single agent's perspective by the following lemma. 

\begin{lemma}\label{lm:aug_err_relation}
Suppose the system given in \eqref{eq:obsv_dynm} is observable and satisfies the Assumption \ref{ass:initialestcov}. Then, given the agent dynamics and the estimation based control \eqref{eq:infinite_input_protocol} with control law as $F$, there exists a affine mapping between the estimation error of the $i^{th}$ agent, ${}^{i}e$, and the augmented MAS estimation error, $e$, as follows: 
\begin{equation}\label{eq:aug_error_relation}
    \begin{split}
        e(t+1) &= H_{i}(t+1){}^{i}e(t+1)+\alpha_{i}(t+1), \ \forall i\in \mathcal{V},\ \forall t \ge 0 
    \end{split}
\end{equation}

\end{lemma}

\emph{Proof}.
The proof can be shown by induction. Let the estimation error at time step $t$ satisfies \eqref{eq:aug_error_relation}. To verify that the \eqref{eq:aug_error_relation} is satisfied at the next time step $t+1$ with the definitions of $H_{i}(t+1)$ and $\alpha_{i}(t+1)$, the dynamics of the estimation error of the $i^{th}$ agent is considered. 
By \eqref{eq:aug_error_relation}, \eqref{eq:agent_p_err_inf} can be stated as follows:
\begin{equation}\label{eq:err_predict_stability}
\begin{split}
    {}^{i}e^{-}(t+1) &= D_{1}{}^{i}e(t)+D_{12}(H_{i}(t){}^{i}e(t)+\alpha_{i}(t))+\tilde{w}(t) \\
    &= {}^{i}D_{2}(t){}^{i}e(t)+D_{12}\alpha_{i}(t)+\tilde{w}(t)
\end{split}
\end{equation}
And the updated estimation error is derived as follows: 
\begin{equation}\label{eq:err_update_stability}
    \begin{split}
   {}^{i}e(t+1)&={}^{i}D_{22}(t+1){}^{i}D_{2}(t){}^{i}e(t)+{}^{i}D_{22}(t+1)D_{12}\alpha_{i}(t)  + \zeta_{i}(t+1)
    \end{split}
\end{equation}
By concatenating \eqref{eq:err_update_stability} for all agents $i\in \mathcal{V}$, the MAS augmented estimation error, $e$, can be formulated as follows:
\begin{equation}
    \begin{split}
        e(t+1) &= \tilde{F}(t+1)e(t)+\gamma(t+1) \\
        &= \tilde{F}(t+1)H_{i}(t){}^{i}e(t)+\tilde{F}(t+1)\alpha_{i}(t) + \gamma(t+1) \\ 
        &= H_{i}(t+1){}^{i}e(t+1)+\alpha_{i}(t+1)
    \end{split}
\end{equation}
Without loss of generality, the matrix ${}^{i}D_{22}(t+1){}^{i}D_{2}(t)$ is invertible as it governs the estimation error dynamics \eqref{eq:err_update_stability} induced by the stochastic linear dynamical system. This completes the proof of Lemma \ref{lm:aug_err_relation}.
\hfill\(\blacksquare\)

Based on the derived affine mapping between $e$ and ${}^{i}e$, we are ready to present the stability of the proposed estimation algorithm. As our paper considers the stochastic MAS, the estimation error can be regarded as a super martingale of the Lyapunov functions, which satisfies the following conditions: 
\begin{equation}\label{eq:stochasticLyapunov}
\left\{\begin{aligned}
&V(e(t),t) = 0 , \ \ \ e(t)=\mathbf{0}\\
&V(e(t),t) > 0 , \ \ \ e(t)\neq\mathbf{0}\\
&V(e(t),t) \rightarrow \infty, \ \ \ e(t)\rightarrow \infty
\end{aligned}\right., \ \ \ \forall t
\end{equation}
\begin{equation}\label{eq:stochasticLyapunov2}
\Delta V(t+1,t) < 0, \ \ \ \forall k
\end{equation}
where $\Delta V(t+1,t):=V(\mathbb{E}\left[e(t+1)|e(t)\right],t+1) - V(e(t),t)$. \\
$\ $

\begin{theorem}
Given the MAS dynamics and the control protocol as in \eqref{eq:infinite_input_protocol}, the proposed distributed estimation algorithm is globally asymptotically stable in the sense of Lyapunov if the system \eqref{eq:obsv_dynm} is observable.  
\end{theorem}

\emph{Proof}.
Let us define the Lyapunov function $V:\mathbb{R}^{nN}\times\mathbb{N}\rightarrow\mathbb{R}$ of the estimation error of the $i^{th}$ agent as follows: 
\begin{equation}\label{eq:estLyapunovfunction}
    \begin{split}
    V({}^{i}e(t),t):=&{}^{i}e^{\mathrm{T}}(t)\left({}^{i}\Sigma(t)\right)^{-1}{}^{i}e(t)\\
    \triangle V(t+1,t):=& V(\mathbb{E}[{}^{i}e(t+1)|{}^{i}e(t)],t+1)-V({}^{i}e(t),t)
    \end{split}
\end{equation}

As ${}^{i}\Sigma(t)$ is positive definite and bounded by lemma \ref{lm:obsv_lm}, There exists $\left({}^{i}\Sigma(t)\right)^{-1} \succ 0$. Therefore, $V$ is a quadratic function which satisfies the condition in \eqref{eq:stochasticLyapunov}. Besides, using \eqref{eq:err_predict_stability}, the conditional expectation $\mathbb{E}[{}^{i}e(t+1)|{}^{i}e(t)]$ is given by 
\begin{equation*}
    \mathbb{E}[{}^{i}e(t+1)|{}^{i}e(t)] = {}^{i}D_{22}(t+1){}^{i}D_{2}(t){}^{i}e(t)
\end{equation*}
Then, by applying equation \eqref{eq:err_update_stability}, $\triangle V$ can be written as follows: 
\begin{equation}
    \begin{split}
        \triangle V(t+1,t) =& -{}^{i}e^{\mathrm{T}}(t)\big(({}^{i}\Sigma(t))^{-1} - {}^{i}D_{2}^{\mathrm{T}}(t){}^{i}D_{22}^{\mathrm{T}}(t+1) \times ({}^{i}\Sigma(t+1))^{-1}{}^{i}D_{22}(t+1){}^{i}D_{2}(t)\big){}^{i}e(t)\\
        =& -{}^{i}e^{\mathrm{T}}(t)\mathcal{M}_{i}(t+1){}^{i}e(t)
    \end{split}
\end{equation}
To satisfies the condition in \eqref{eq:stochasticLyapunov2}, $\mathcal{M}_{i}(t+1)$ should be a positive definite matrix. By applying Lemma \ref{lm:aug_err_relation}, the predicted estimation error covariance of the $i^{th}$ agent is derived using \eqref{eq:err_predict_stability} as:
\begin{equation}\label{eq:predicted_sig_stab}
      {}^{i}\Sigma^{-}(t+1)= {}^{i}D_{2}(t){}^{i}\Sigma(t){}^{i}D_{2}^{\mathrm{T}}(t)+D_{12}\eta_{i}(t)D_{12}^{\mathrm{T}}+\Sigma_{\tilde{w}}
\end{equation}
Correspondingly, the updated estimation error covariance can be computed as follows:
\begin{equation}\label{eq:updated_sig_stab}
\begin{split}
    {}^{i}\Sigma(t+1) =& {}^{i}D_{22}(t+1){}^{i}\Sigma^{-}(t+1)\\
    =& {}^{i}\Sigma^{-}(t+1)-{}^{i}\Sigma^{-}(t+1)C_{i}^{\mathrm{T}} \big(C_{i}{}^{i}\Sigma^{-}(t+1)C_{i}^{\mathrm{T}} + C_{i}\mathrm{\Xi}_{i}C_{i}^{\mathrm{T}}\big)^{-1}C_{i}{}^{i}\Sigma^{-}(t+1)
\end{split}
\end{equation}
and using the definition of ${}^{i}D_{22}$, ${}^{i}\Sigma(t+1)$ can be defined as: 
\begin{equation}\label{eq:updatecov_predictcov}
    {}^{i}\Sigma(t+1) = {}^{i}D_{22}(t+1){}^{i}\Sigma^{-1}(t+1)
\end{equation}
By applying the matrix inversion lemma, \eqref{eq:updated_sig_stab} is rewritten as: 
\begin{equation}\label{eq:updated_sig_stab2}
    \begin{split}
    {}^{i}\Sigma(t+1)=& \big( ({}^{i}\Sigma^{-}(t+1))^{-1}+C_{i}^{\mathrm{T}}( C_{i}\mathrm{\Xi_{i}}C_{i}^{\mathrm{T}})^{-1}C_{i}\big)^{-1}
    \end{split}
\end{equation}

And using \eqref{eq:updated_sig_stab2}, multiplying ${}^{i}\Sigma(t+1)$ to the left and the right sides of $\big({}^{i}\Sigma(t+1)\big)^{-1}$ and applying \eqref{eq:updatecov_predictcov} yields:
\begin{equation}
    {}^{i}\Sigma(t+1) = {}^{i}D_{22}(t+1)\big({}^{i}\Sigma^{-1}(t+1) +\mathcal{W}_{i}(t+1)\big){}^{i}D_{22}^{\mathrm{T}}(t+1)
\end{equation}
where $ \mathcal{W}_{i}(t+1)={}^{i}\Sigma^{-}(t+1)C_{i}^{\mathrm{T}}(C_{i}\mathrm{\Xi_{i}}C_{i}^{\mathrm{T}})^{-1}C_{i}{}^{i}\Sigma^{-}(t+1)$. It is trivial to show  that matrix $\mathcal{W}_{i}\succ 0, \  \forall t$. Recalling ${}^{i}\Sigma^{-}(t+1)$ denoted in \eqref{eq:predicted_sig_stab}, $\big({}^{i}\Sigma(t+1)\big)^{-1}$ can be rewritten by: 
\begin{equation}\label{eq:inv_cov}
    \begin{split}
       \big({}^{i}\Sigma(t+1)\big)^{-1}=& \big({}^{i}D_{22}^{\mathrm{T}}(t+1)\big)^{-1}\big({}^{i}D_{2}(t){}^{i}\Sigma(t){}^{i}D_{2}^{\mathrm{T}}+\Sigma_{\tilde{w}}+D_{12}\eta_{i}(t)D_{12}^{\mathrm{T}}+\mathcal{W}_{i}(t+1)\big)^{-1}\big({}^{i}D_{22}(t+1)\big)^{-1}
    \end{split}
\end{equation}
By applying \eqref{eq:inv_cov}, $\mathcal{M}_{i}(t+1)$ can be redefined as:
\begin{equation}\label{eq:mispd1}
\begin{split}
\mathcal{M}_{i}(t+1)=&({}^{i}\Sigma(t))^{-1}-{}^{i}D_{2}^{\mathrm{T}}(t)\big({}^{i}D_{2}(t){}^{i}\Sigma(t){}^{i}D_{2}^{\mathrm{T}}(t)+D_{12}\eta_{i}(t)D_{12}^{\mathrm{T}}+\Sigma_{\tilde{w}}+\mathcal{W}_{i}(t+1)\big)^{-1}{}^{i}D_{2}(t)
\end{split}
\end{equation}
After going through tedious conversion using the matrix inversion lemma, \eqref{eq:mispd1} can be rewritten as follows: 
\begin{equation}\label{eq:mispd2}
\begin{split}
    \mathcal{M}_{i}(t+1)=&
    ({}^{i}\Sigma(t))^{-1}
    \big(({}^{i}\Sigma(t))^{-1}+{}^{i}D_{2}^{\mathrm{T}}(t)(D_{12}\eta_{i}(t)D_{12}^{\mathrm{T}}+\Sigma_{\tilde{w}}+\mathcal{W}_{i}(t+1))^{-1}{}^{i}D_{2}(t)\big)^{-1}({}^{i}\Sigma(t))^{-1}
\end{split}
\end{equation}
As $({}^{i}\Sigma(t))^{-1} \succ 0$ and ${}^{i}D_{2}^{\mathrm{T}}(t)(D_{12}\eta_{i}(t)D_{12}^{\mathrm{T}}+\Sigma_{\tilde{w}}+\mathcal{W}_{i}(t+1) \succ 0$ in \eqref{eq:mispd2}, one can verify $\mathcal{M}_{i}(t+1)\succ 0,\ \forall t$, which guarantees the Lyapunov function satisfies \eqref{eq:stochasticLyapunov} and \eqref{eq:stochasticLyapunov2}. This proves that the estimation error is globally asymptotically stable.
\hfill\(\blacksquare\)

\section{Numerical Simulation} \label{sc:simulation}
\begin{figure}[htpb]
\centering
\includegraphics[width=0.8\textwidth]{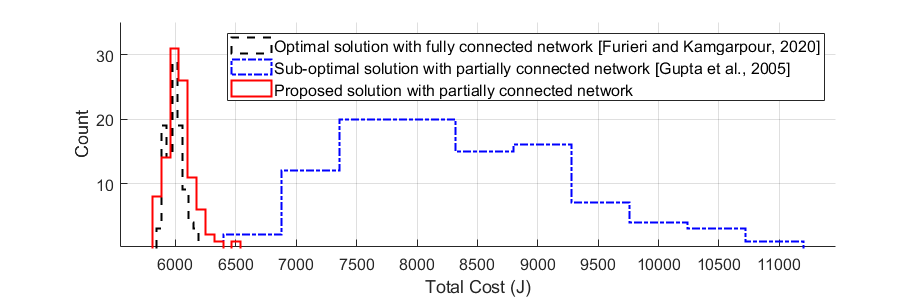}
\caption{Cost value statistics histogram (Monte Carlo simulations
with 100 runs).}\label{fig:finite_cost_hist}
\end{figure}

The effectiveness of the proposed algorithm is demonstrated with an illustrative MAS example. The MAS consists of five agents whose dynamics and objective are specified by the following parameter sets: $A=1,\ B=1,\ \Theta_{i}(t)=1$,\ ${}^{i}\mathrm{\Xi}(t)=diag(1, 1, 1, 2, 1), \forall i\in\mathcal{V}$, $\forall t\in\{0,\cdots,T\}$,\ $T=5$,\ $\mathcal{Q} = I_{6}\otimes (5I_{5} - 1_{5\times 5})$, and $\mathcal{R} = I_{25}$. The MAS network topology is set to be partially connected, the same as the one in \citep{kwon2018sensing}. To validate the performance of the proposed algorithm, we conduct a comparative analysis with two different scenarios: i) MAS with the fully connected network, which is free from network topological constraint; and ii) MAS with the same (partially connected) network topology where non-neighboring agent information is not available to each agent. For the second case, we test the suboptimal method presented in \citep{gupta2005sub}. The simulation results are shown in Figure. \ref{fig:finite_cost_hist}. By virtue of the virtual interactions between non-neighboring agents, our algorithm outperforms the existing method in the partially connected network, and even matches the fully connected network case despite the network topological constraints.

\begin{figure}[htpb]
\centering
\includegraphics[width=0.8\textwidth]{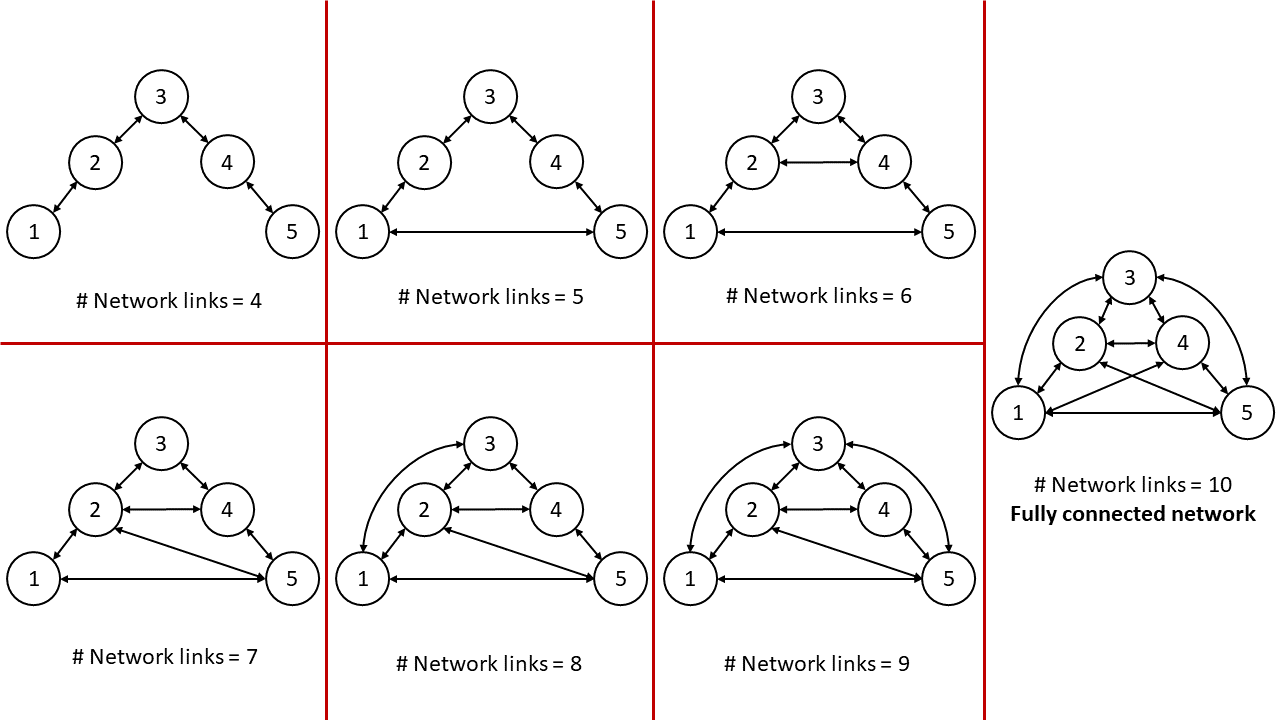}
\caption{Different network topology with 5 agents.}\label{fig:networkconfig}
\end{figure}

\begin{figure}[htpb]
\centering
\includegraphics[width=0.7\textwidth]{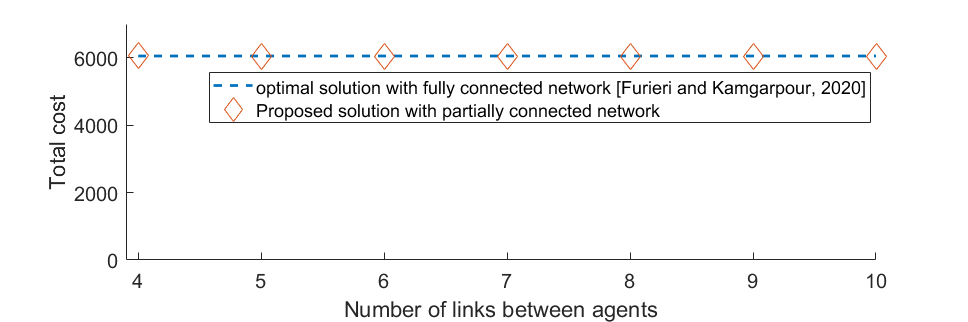}
\caption{Cost value comparison with a different number of network links between agents.}\label{fig:networkconnection_5agents}
\end{figure}

To further demonstrate the performance of the proposed algorithm, we have performed additional simulations with respect to the different number of "real" interactions. It is worth noting that the network links between agents yield the “real" interaction while those with no explicit links create the virtual interaction. MAS with five agents has been simulated under different network topology shown in the below Figure. \ref{fig:networkconfig}. 
All experiments followed the same setting except the network topology. Apparently, the number of virtual interactions decreases as the number of network links between agents increase whereby we can analyze the effect of ratio between real and virtual interactions. The performance of our proposed method under different network settings is depicted in the below Figure. \ref{fig:networkconnection_5agents}.
The result shows that the performance of our proposed method does not vary much with respect to the ratio between virtual to real interactions. At the cost of some onboard computational resources to estimate non-neighboring agents, our proposed method provides the clear advantage of having optimal performance with fewer network connections. 

\section{Conclusions}\label{sc:conclusion}
This paper has proposed a novel design procedure of the optimal distributed control for the linear stochastic MAS, generally subject to network topological constraints. The proposed method gets around the network topological constraint by employing the distributed estimator, whereby each agent can exploit the non-neighboring agent's information. Future work will include the theoretical performance guarantee of the proposed distributed control-estimation synthesis such as stability analysis, and a further extension to the infinite time horizon case for practical use.

\bibliographystyle{unsrtnat}
\bibliography{template}  

\end{document}